# SUPPLY-POWER-CONSTRAINED CABLE CAPACITY MAXIMIZATION USING DEEP NEURAL NETWORKS

*Junho Cho[1*], Sethumadhavan Chandrasekhar[1], Erixhen Sula[2], Samuel Olsson[3], Ellsworth Burrows[1], Greg Raybon[1], Roland Ryf[1], Nicolas Fontaine[1], Jean-Christophe Antona[4], Steve Grubb[5], Peter Winzer[1], and Andrew Chraplyvy[1]*

[1]*Nokia Bell Labs, Holmdel, NJ, USA*, [2]*EPFL, Lausanne, Switzerland*, [3]*Nokia Corporation, Murray Hill, NJ, USA*,
[4]*Alcatel Submarine Networks, Nozay, France*, [5]*Facebook, Menlo Park, CA, USA*
*email: junho.cho@nokia-bell-labs.com

## Abstract

We experimentally achieve a 19% capacity gain per Watt of electrical supply power in a 12-span link by eliminating gain flattening filters and optimizing launch powers using machine learning by deep neural networks in a massively parallel fiber context.

## 1 Introduction

Massive spatial parallelism has been shown to maximize the capacity and to minimize the cost/bit of submarine optical cables, in view of a constrained electrical supply power per cable [1]–[3]. Owing to the resulting optical power dilution among many parallel fibers, transmission is pushed from nonlinearly-optimum launch powers to the linear regime. The logarithmically reduced spectral efficiency from a lower delivered optical signal-to-noise ratio (OSNR) per fiber is then linearly over-compensated by the increased spatial multiplicity of the cable to yield a higher total cable capacity [1], [2]. The capacity $C$ per Watt of electrical supply power $\mathcal{P}_E$, both per spatial path, becomes a key figure of merit in such systems [1],

$$m = C/\mathcal{P}_E. \qquad (1)$$

This new submarine cable design space asks for revisiting such fundamental topics as *(i)* the need for gain-flattening filters (GFFs) in conjunction with optical amplifiers (as GFFs, which are universally used in all submarine systems today, are lossy and hence waste precious cable supply power, thus potentially reducing $m$), and *(ii)* the optimum optical channel power allocation strategy (in particular since capacity-optimizing water-filling techniques used in wireless communications are known to bring a greatest gain in the low-SNR regime). This paper addresses both topics and, on an exemplary 12-span 744-km straight-line system, experimentally achieves a capacity gain per Watt of electrical supply power of up to 19%. Higher gains in $m$ are expected for longer links and for pump-sharing architectures across amplifier arrays.

Accurately predicting the signal and noise power evolution of a long chain of un-flattened optical amplifiers for arbitrary transmit (TX) power profiles is difficult, as a small change in the TX power spectral density (PSD) or in the spectral link characteristics may cause a complicated evolution of signal and noise powers through the system, making it intractable to computationally solve the problem using analytical or numerical physics-based optical amplifier models. We therefore resort to machine learning [4] and build a deep neural network (DNN) as a digital twin of our optical fiber link. Once properly trained with experimental link data, the DNN allows for an off-line gradient-descent (GD) optimization whose optimized results are then verified experimentally.

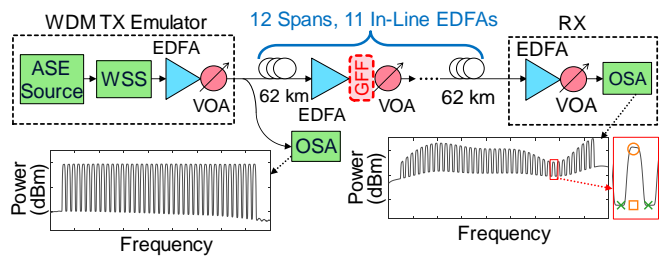

Fig. 1. Experimental setup. *Inset:* Measured optical spectrum.

## 2 Experimental Methodology and Setup

Massively parallel submarine cables will operate at low-enough optical signal powers to neglect fiber nonlinearities [2], and probabilistic constellation shaping allows to finely adapt each wavelength channel's transponder to the specific SNR of that channel [5]–[8]. This lets measurements of the delivered OSNR be a good basis for estimating polarization- and wavelength-division multiplexed (WDM) system capacities as

$$C = 2R_s \sum_{k=1}^{K} \log_2(1 + \eta\, SNR_k), \qquad (2)$$

where $SNR_k$ is the OSNR of the $k$-th of $K$ WDM channels (normalized to one polarization and a reference bandwidth equal to the symbol rate $R_s$), and $\eta \leq 1$ accounts for transponder implementation penalties. We use $\eta = 1$ without limiting the generality of the optimization methodology.

In order to determine $SNR_k$, we use the WDM channel emulation method shown in Fig. 1: Amplified spontaneous emission (ASE) from an Erbium-doped fiber amplifier (EDFA) is filtered by a wavelength selective switch (WSS) to generate 40 slots of 50-GHz bandwidth ASE (emulating 40 signal channels, as customarily used for WDM loading channels [9], [10]), interleaved with 39 "empty" 50-GHz slots. At any point within the system, each $SNR_k$ can then be estimated by an optical spectrum analyzer (OSA) taking the ratio of emulated signal power $S_k$ to ASE power $N_k$, interpolated between two empty slots, cf. rightmost inset to Fig. 1. (In practice, due to the finite WSS extinction ratio, the TX SNR is limited to ~45 dB, which does not constrain our measurements due to the much higher in-line noise added by the link.) The emulated WDM channels at the WSS output are boosted by a TX EDFA and attenuated by a variable optical attenuator (VOA) to produce a set of desired optical launch powers $\boldsymbol{P}_{1:40}$; we use the notation $\boldsymbol{X}_{1:K} \coloneqq [X_1, \dots, X_K]$ throughout the paper. An example flat TX signal power allocation across a system bandwidth of 4 THz is shown in the leftmost inset to Fig. 1, together with the received (RX) PSD produced after 12 spans of 62-km Corning® Vascade® EX3000 fiber with 0.16-dB/km loss, padded by VOAs to a span loss of ~16.5 dB in order to operate our 744-km straight-line system in a lower-OSNR regime pertinent to the targeted massively parallel submarine application [1], [2]. (We also performed loop experiments without VOA-padded spans, but these were limited by artifacts from GFF-free loop operation, even with lumped loop equalization.) Since launch powers are low and fiber nonlinearities are negligible (as quantified below), padding the spans at the beginning of a span is equally permissible as padding them at the end. Each span is followed by a custom-designed single-stage EDFA with a removable GFF, as we want to compare the capacity gain by removing GFFs in an otherwise identical system.

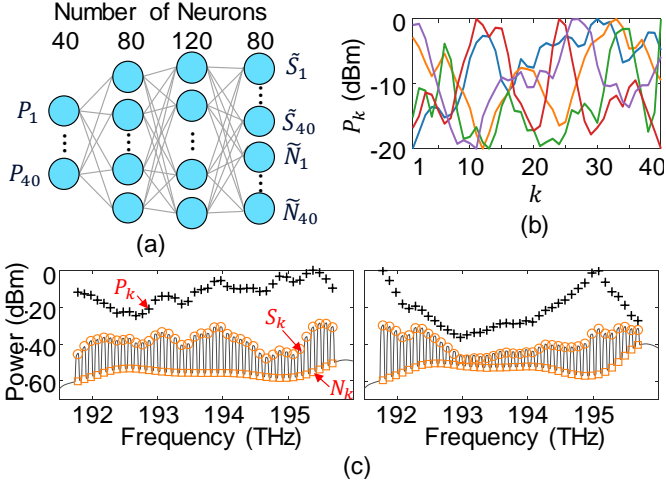

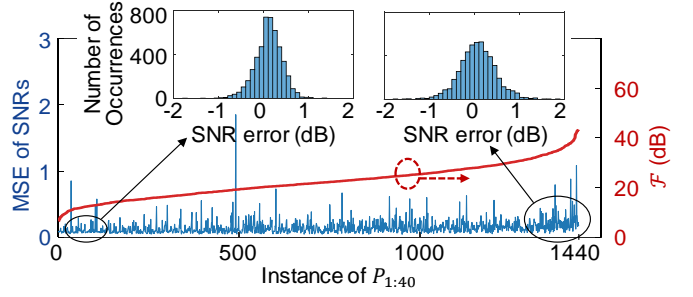

Fig. 3. Accuracy of SNR prediction of DNN.

Fig. 2. (a) Structure of the DNN, (b) five random TX power profiles $\boldsymbol{P}_{1:40}$ with $\mathcal{F} = 20$ dB, and (c) two examples of TX powers (pluses), measured RX PSDs (solid lines), and DNN-predicted signal (circles) and noise (squares).

## 3 Training the DNN with Experimental Data

Since each of the signal and noise powers $(\boldsymbol{S}_{1:40}, \boldsymbol{N}_{1:40})$ at the output of the link depends on the full set of launch powers $\boldsymbol{P}_{1:40}$ in a way that is difficult to accurately model based on amplifier physics, we resort to machine learning and construct a DNN as a digital twin of our experimental link. As shown in Fig. 2(a), our DNN has 40 input neurons ($\boldsymbol{P}_{1:40}$), two hidden layers with 80 and 120 neurons each, and 80 output neurons for the predicted signal and noise powers ($\tilde{\boldsymbol{S}}_{1:40}, \tilde{\boldsymbol{N}}_{1:40}$) at the output of the link. Linear, sigmoid, and softplus activation functions [11] are used. Numbers of neurons and activation functions are chosen to minimize the mean squared error (MSE) of measurement ($\boldsymbol{S}_{1:40}, \boldsymbol{N}_{1:40}$) and prediction ($\tilde{\boldsymbol{S}}_{1:40}, \tilde{\boldsymbol{N}}_{40}$).

Each DNN training process starts by configuring one of two link setups (i.e., with and without GFFs) and choosing one of three total available electrical supply power levels $\mathcal{P}_E$ (we consider only *electrical pump powers* and ignore less fundamental overheads from amplifier control). The overall electrical power is spread approximately evenly across all 11 in-line EDFAs such that the optical output power summed over all 40 signal channels ($\mathcal{P}_O$) is equal for every EDFA. The TX VOA is adjusted to provide the same total TX power $\mathcal{P}_O$ during this process. The EDFAs, when operated with GFFs, have a gain ripple < 1.5 dB across the studied 4-THz amplification band for all chosen operating conditions and operate at wall-plug efficiencies of 2.7, 6.6, 8.2% (measured after the GFF), respectively for $\mathcal{P}_E$ =1.09, 2.27, 7.53 W. Next, we measure $\boldsymbol{S}_{1:40}$ and $\boldsymbol{N}_{1:40}$ for 1440 randomly generated $\boldsymbol{P}_{1:40}$ subject to $\sum_{k=1}^{40} P_k = \mathcal{P}_O$, and with a peak-to-peak channel power excursion $\mathcal{F} = \max_{i,j}(|P_i - P_j|)$ that we gradually increase from 6 dB to 45 dB; 5 representative instances of TX signal powers with 20-dB peak-to-peak excursion are depicted in Fig. 2(b). We avoid implausibly fast changes of $P_k$ over a narrow frequency range by applying a moving average to each TX power profile. We also ensured that the 1440 random power profiles uniformly fill the frequency-power rectangle by maximizing the relative entropies between the power profiles normalized to sum to one. Of the 1440 recorded data sets, 90% are used for training and 10% for validation of the DNN. For all 6 test cases, the DNN rapidly converges with a minuscule MSE difference between training and validation sets (indicating the absence of overfitting [12]). Figure 2(c) shows two representative examples of $\boldsymbol{P}_{1:40}$(pluses), and the corresponding measured RX PSDs (solid lines) for $\mathcal{P}_E = 2.27$ W without GFFs. Also shown are the DNN-predicted $\tilde{\boldsymbol{S}}_{1:40}$, $\tilde{\boldsymbol{N}}_{1:40}$ (circles and squares), with great agreement between measurement and prediction. Figure 3 shows that the DNN predicts the channel SNRs for 1440 random $\boldsymbol{P}_{1:40}$ with very small prediction errors, for the test case of $\mathcal{P}_E = 2.27$ W without GFFs and across a wide range of $\mathcal{F}$. The small prediction error justifies the optimization of signal power allocations based on DNNs.

## 4 Capacity Maximization and Verification

We next perform gradient descent (GD) capacity maximization off-line based on Eq. (2) and the trained DNN, cf. Fig. 4(a). The result is a capacity-maximizing TX power profile $\boldsymbol{P}_{1:40}$. Figure 4(b) shows, for $\mathcal{P}_E = 2.27$ W without GFFs, three example capacity optimizations, one starting from a flat $\boldsymbol{P}_{1:40}$ (blue) and the other two from initial conditions with poorer capacity. Figure 4(c) shows initial (blue crosses) and converged (orange dots) capacities for all of the 1440 randomly chosen power profiles with varying $\mathcal{F}$ (red dots), revealing reliable GD-DNN convergence to the same optimal capacity in almost all cases even in the absence of GFFs, indicating that the capacity surface is close to concave within the boundaries of our experimental conditions, hence there may be *only one optimal* TX power allocation. Initial (left) and converged (right) SNRs for all 1440 realizations are shown in Fig. 4(d) across the system bandwidth. Remarkably, the converged SNR distribution is reasonably flat to within 4 dB in most cases, with minimal capacity variations between these converged solutions. The capacity of a completely flat SNR is 25.6 Tb/s, which is close to the experimental optimum of 25.9 Tb/s but further off the capacities of a flat TX signal power profile (24.8 Tb/s) and a flat RX signal power profile (24.5 Tb/s), in contrast to the findings of Ref. [13] for conventional systems using GFFs. Importantly, our approach does *not* subjectively favor any capacity optimization strategy based on possibly misguiding intuition (e.g., "flat RX signal power profile", "flat TX signal power profile", "flat SNR", or "water-filling"), but objectively optimizes the TX signal powers solely by following the gradient trajectory of ascending system capacity. Also, note that an *experimental* GD solution is uniquely enabled by our DNN approach, as estimating only a single gradient in our case requires 41 measurements of 4-THz RX PSDs. In our fully automated system, we are able to measure 180 RX PSDs per hour, hence it would require >11 years to perform a full GD optimization for 1440 TX power profiles with 300 GD iterations! On the other hand, the optimization process takes only 9 hours using the DNN approach. This >10,000× speed-up reveals the power of machine learning, enabling us to experimentally determine a multi-dimensional capacity surface, which would have been impossible by physical experiments alone.

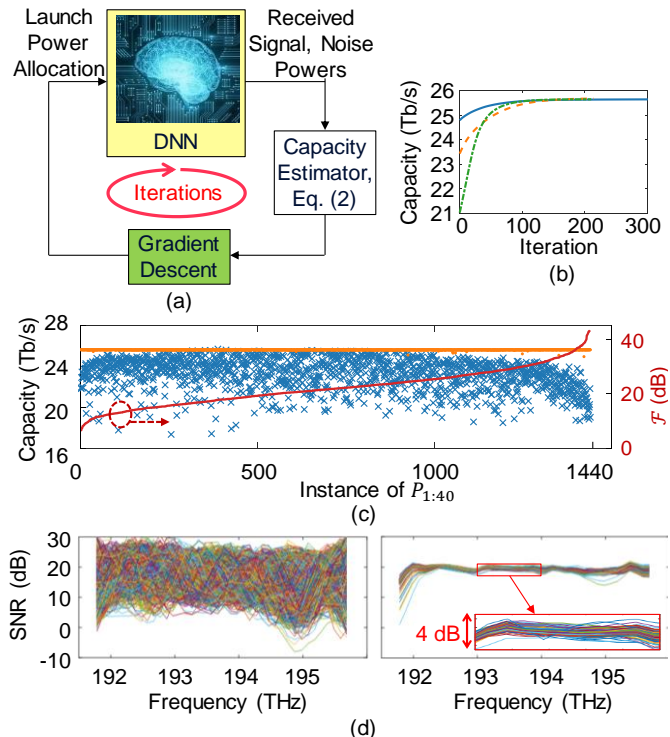

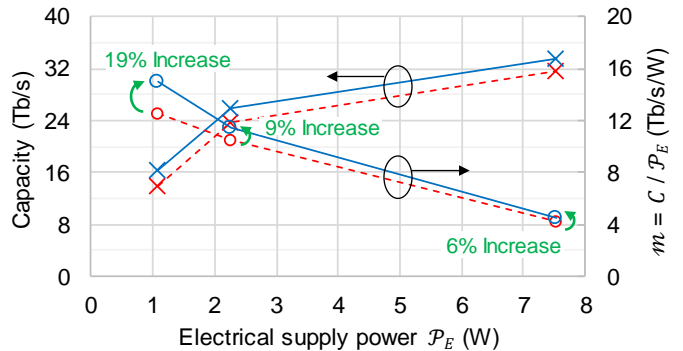

Fig. 5. Optimized $C$ (crosses) and $m$ (circles) as a function of the total electric pump power in systems with (dashed) and without (solid) GFFs.

Fig. 4. (a) Capacity optimization using DNN and GD, (b) typical convergence of the GD, (c) capacity of the initial (blue crosses) and converged (orange dots) TX power profiles, (d) RX SNRs of the 1440 random TX power profiles at iteration 0 (left) and after convergence (right).

As a last step, we validate the results of the DNN-based GD optimization by loading the optimized TX power profiles into the experimental WDM emulator and measuring the resulting RX power profile. The capacity predicted by the DNN is within a 1.1% error of the experimentally measured capacity in all test cases. We also verify that the maximum channel power occurring anywhere within the system for the optimized power profile is below -4 dBm per 50 GHz slot for $\mathcal{P}_E = 1.09$ W, both with and without GFFs. This justifies the initial assumption of neglecting fiber Kerr nonlinearities, as it operates 7 dB below a deployed cable that shows nonlinear peak performance at 3 dBm per 50-GHz channel [10].

Figure 5 shows the actually measured capacity $C$ (left axis) and the figure of merit $m$ (right axis). Dashed lines represent systems with GFFs and solid lines without GFFs, all with optimized TX power allocations. The experimental results show that: *(i)* systems without GFFs achieve a better power efficiency than systems with GFFs, *(ii)* $m$ increases with decreasing $\mathcal{P}_E$ until the EDFA pump current approaches the pump's lasing threshold, even at a significantly reduced EDFA wall-plug efficiency of only 2.7% at that operating point; hence, operating the pumps at higher power (and hence at higher efficiencies) and sharing them across multiple EDFAs will further increase $m$; and *(iii)* when the system operates at maximum efficiency (at largest $m$), both $C$ and $m$ can be increased by 19% by eliminating GFFs from the system.

## 5 Conclusion

We used experimental signal and noise data from a 12-span 744-km straight-line EDFA link to train a DNN as a digital twin of the experimental system. The DNN accurately predicts received signal and noise powers for arbitrary transmit signal power distributions, even without GFFs as part of the link. A gradient descent based transmit power profile optimization performed on the DNN is about 10000 times faster than what would be possible using measurements alone and objectively predicts optimized launch power profiles. In the context of a massively parallel electrical supply power constrained system (such as a submarine optical cable), we demonstrate a 19% improvement in achievable cable capacity.

## 6 Acknowledgements

We acknowledge Amonics Corp for designing and supplying the EDFAs with removable GFFs, and Corning for the loan of the EX3000 fiber used in this experiment.

# 7 References

[1] O. V. Sinkin, A. V. Turukhin, Y. Sun, et al., "SDM for power-efficient undersea transmission," J. Lightw. Technol., 2017, **36,** (2), pp. 361–371.

[2] R. Dar, P. J. Winzer, A. R. Chraplyvy, et al., "Cost-optimized submarine cables using massive spatial parallelism," J. Lightw. Technol., 2018, **36**, (18), pp. 3855–3865.

[3] P. Pecci and O. Courtois, "SDM: A revolution for the submarine industry," Submarine Telecoms Forum, 2019, **106**, pp. 38–41.

[4] D. Zibar, M. Piels, R. Jone., et al., "Machine learning techniques in optical communication," in J. Lightw. Technol., 2016, **34**, (6), pp. 1442–1452.

[5] G. Böcherer, F. Steiner, and P. Schulte., "Bandwidth efficient and rate-matched low-density parity-check coded modulation," IEEE Trans. Commun., 2015, **63**, (12), pp. 4651–4665.

[6] F. Buchali, F. Steiner, G. Böcherer, et al., "Rate adaptation and reach increase by probabilistically shaped 64-QAM: An experimental demonstration," J. Lightw. Technol., 2016, **34**, (7), pp. 1599–1609.

[7] A. Ghazisaeidi, I. F. d. J. Ruiz, R. R.-Muller, et al., "65 Tb/s transoceanic transmission using probabilistically-shaped PDM-64QAM," Proc. Eur. Conf. Opt. Commun., Dusseldorf, Germany, Sep. 2016, Paper Th.3.C.4.

[8] J. Cho and P. J. Winzer, "Probabilistic constellation shaping for optical fiber communications," J. Lightw. Technol., 2019, **37**, (6), pp. 1590–1607.

[9] D. J. Elson, L. Galdino, R. Maher, et al., "High spectral density transmission emulation using amplified spontaneous emission noise," Opt. Lett., 2016, **41**, (1), pp. 68–71.

[10] J. Cho, X. Chen, S. Chandrasekhar, et al., "Trans-Atlantic field trial using high spectral efficiency probabilistically shaped 64-QAM and single-carrier real-time 250-Gb/s 16-QAM," J. Lightw. Technol., 2018, **36**, (1), pp. 103–113.

[11] P. Ramachandran, Z. Barret, and Q. V. Le, "Searching for activation functions," arXiv preprint, 2017, arXiv:1710.05941.

[12] N. Srivastava et al., "Dropout: a simple way to prevent neural networks from overfitting," J. Mach. Learn. Res., 2017, **15**, (1), pp. 1929–1958.

[13] A. Kam, P. Mehta, D. Evans, et al., "Pre-emphasis-based equalization strategy to maximize cable performance," Proc. SubOptic, New Orleans, LA, USA, Apr. 2019, Paper OP14-1.